# Cloning of Dirac fermions in graphene superlattices

L. A. Ponomarenko[1], R. V. Gorbachev[2*], G. L. Yu[1], D. C. Elias[1], R. Jalil[2], A. A. Patel[3], A. Mishchenko[1], A. S. Mayorov[1], C. R. Woods[1], J. R. Wallbank[3], M. Mucha-Kruczynski[3], B. A. Piot[4], M. Potemski[4], I. V. Grigorieva[1], K. S. Novoselov[1], F. Guinea[5], V. I. Fal'ko[3], A. K. Geim[1,2]

[1] School of Physics & Astronomy, University of Manchester, Manchester M13 9PL, UK
[2] Centre for Mesoscience & Nanotechnology, University of Manchester, Manchester M13 9PL, UK
[3] Physics Department, Lancaster University, LA1 4YB, UK
[4] Laboratoire National des Champs Magnetiques Intenses, CNRS-UJF-UPS-INSA, F-38042 Grenoble, France
[5] Instituto de Ciencia de Materiales de Madrid (CSIC), Sor Juana Inés de la Cruz 3, Madrid 28049, Spain

**Lateral superlattices have attracted major interest as this may allow one to modify spectra of two dimensional (2D) electron systems and, ultimately, create materials with tailored electronic properties[1-8]. Previously, it proved difficult to realize superlattices with sufficiently short periodicity and weak disorder, and most of the observed features could be explained in terms of commensurate cyclotron orbits[1-4]. Evidence for the formation of superlattice minibands (so called Hofstadter's butterfly[9]) has been limited to the observation of new low-field oscillations[5] and an internal structure within Landau levels[6-8]. Here we report transport properties of graphene placed on a boron nitride substrate and accurately aligned along its crystallographic directions. The substrate's moiré potential[10-12] leads to profound changes in graphene's electronic spectrum. Second-generation Dirac points[13-22] appear as pronounced peaks in resistivity accompanied by reversal of the Hall effect. The latter indicates that the sign of the effective mass changes within graphene's conduction and valence bands. Quantizing magnetic fields lead to Zak-type cloning[23] of the third generation of Dirac points that are observed as numerous neutrality points in fields where a unit fraction of the flux quantum pierces the superlattice unit cell. Graphene superlattices open a venue to study the rich physics expected for incommensurable quantum systems[7-9,22-24] and illustrate the possibility to controllably modify electronic spectra of 2D atomic crystals by using their crystallographic alignment within van der Waals heterostuctures[25].**

Since the first observation of Weiss oscillations[1,2], 2D electronic systems subjected to a periodic potential have been studied in great detail[3-8]. The advent of graphene has rapidly sparked interest in its superlattices, too[13-22]. The principal novelty in this case is the Dirac-like spectrum and the fact that charge carriers are not buried deep under the surface, allowing a relatively strong superlattice potential on a true nanometer scale. One promising avenue for making nanoscale graphene superlattices is the use of a potential induced by another crystal. For example, graphene placed on top of graphite or hexagonal boron nitride (hBN) exhibits a moiré pattern[10-12,26], and graphene's tunneling density of states becomes strongly modified[12,26] indicating the formation of superlattice minibands. The spectral reconstruction occurs near the edges of superlattice's Brillouin zone (SBZ) that are characterized by wavevector $G = 4\pi/\sqrt{3}D$ and energy $E_S = \hbar v_F G/2$ ($D$ is the superlattice period and $v_F$ graphene's Fermi velocity)[12,22].

To observe moiré minibands in transport properties, graphene has to be doped so that the Fermi energy reaches the reconstructed part of the spectrum. This imposes severe constraints on the misalignment angle $\theta$ of graphene relatively to hBN. Indeed, $D$ is given by $\theta$ and the 1.8% difference between the two lattice constants[12]. In the case of perfect alignment ($\theta = 0$), $D$ acquires a maximum value of $\approx 13\pm1$ nm[12], which yields $E_S \approx 0.2$eV. This energy scale corresponds to carrier density $n \approx 3\times10^{12}$ cm$^{-2}$, achievable by field effect doping. However, misorientation by only 2° decreases $D$ twice[12], and 4 times higher $n$ are necessary to reach the SBZ. In practice, studies of the superlattice spectrum in monolayer graphene require $\theta \leq 1°$ (see Methods).

Here we study high-mobility encapsulated graphene devices, similar to those reported previously[27] but a principally new element is added. It is a crystallographic alignment between graphene and hBN with accuracy $\approx 1°$. Figure 1 shows typical behavior of longitudinal and Hall resistivities ($\rho_{xx}$ and $\rho_{xy}$ respectively) for our aligned devices. There is the standard peak in $\rho_{xx}$ at zero $n$, graphene's main neutrality point (NP). In addition, two other peaks appear symmetrically at high doping $n = \pm n_S$. At low temperatures ($T$), the secondary peak at the hole side is stronger than that at the main NP whereas the electron-side peak is ~10 times weaker. The sign reversal of $\rho_{xy}$ (Fig. 1b) cannot be explained by additional scattering and proves that hole(electron)-like carriers appear in the conduction(valence) band of graphene. We attribute the extra NPs to the superlattice potential induced by hBN, which results in minibands featuring isolated secondary Dirac points (DPs) (inset of Fig. 1a). This interpretation agrees with theory[12-22] and the tunneling features reported in ref. 12, including the fact that those were stronger in the valence band.

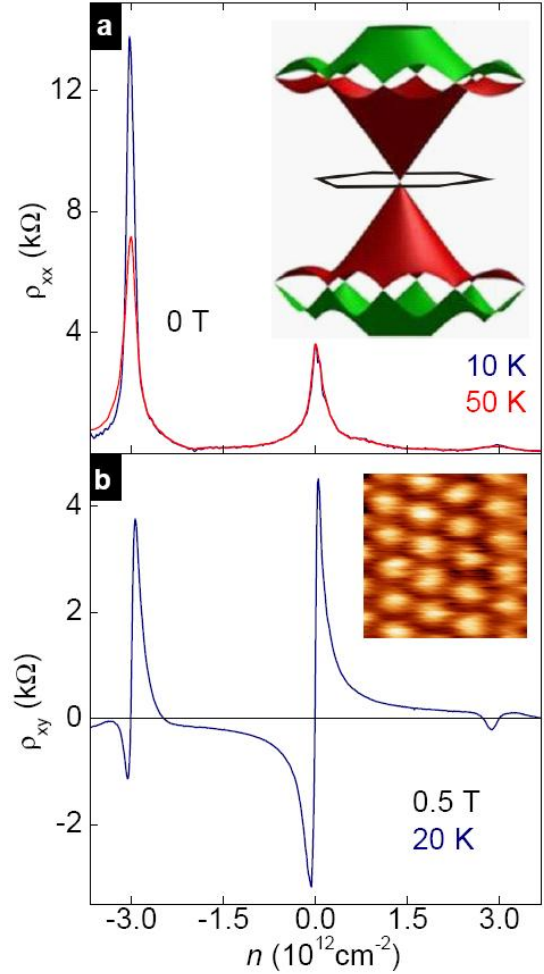

FIG. 1. **Transport properties of Dirac fermions in moiré superlattices. a** – $\rho_{xx}$ as a function of $n$. Positive and negative $n$ correspond to electrons and holes, respectively. The hole-side NP exhibits a strong $T$ dependence (section 1 of Supplementary Info). **b** – $\rho_{xy}$ changes its sign at high electron and hole doping, revealing well-isolated secondary Dirac points. The data are for device A with $n_S \approx 3.0 \times 10^{12}$ cm$^{-2}$, yielding $D \approx 12$nm. We fabricated 11 aligned devices, and 6 of them exhibited essentially the same behavior as shown in the figure, and only $n_S$ varied between 3.0 and 3.8 $\times 10^{12}$cm$^{-2}$ (corresponds to $D \approx 11\pm1$nm). One other device showed $n_S \approx 7.1 \times 10^{12}$cm$^{-2}$ which implies[12] $\theta \approx 1.2°$ and required gate voltages close to dielectric breakdown. Inset in (a): One of possible scenarios for the reconstruction of graphene's spectrum[22]. The band structure is plotted only for the 1st and 2nd SBZ. Secondary Dirac cones appear in both conduction and valence bands at the edges of the SBZ shown by the black hexagon. Where the cones merge, van Hove singularities appear in the density of states (for details, see[22]). Inset in (b): Conductive AFM image of the moiré pattern for one of our devices. The spatial scale is given by the separation of ≈11nm between white spots. Note that, in measurements of $\rho_{xy}$, both positive and negative $B$ were always employed to symmetrize data and subtract even a small contribution due to $\rho_{xx}$.

Near the main NP, the aligned devices exhibit transport characteristics typical for graphene on hBN[27,28]. Conductivity $\sigma(n) = 1/\rho_{xx}$ varies linearly with $n$ and, therefore, can be described by constant mobility $\mu$. For the reported devices, we find $\mu \approx 20–80\times10^3$ cm$^2$V$^{-1}$s$^{-1}$ for $|n| > 10^{11}$cm$^{-2}$. Around the secondary NPs, $\sigma$ depends linearly on ($n-n_S$). The hole-side secondary NP (hSNP) exhibits low-$T$ $\mu$ practically the same as the main NP, whereas near the electron-side secondary NP we find even higher $\mu \approx 30–100\times10^3$ cm$^2$V$^{-1}$s$^{-1}$. However, the main and secondary NPs exhibit very different $T$ dependences of both $\mu$ and minimum conductivities. This is discussed in Supplementary Info and here we only note that the observed $\sigma(T)$ do not support the idea of significant energy gaps induced by the superlattice[19-22]. Furthermore, following the approach described in ref. 29, we have analyzed thermal broadening of the peaks in $\rho_{xx}$ (section 2 of Supplementary Info). The analysis proves that the spectrum at the secondary NPs is linear, that is, Dirac-like, in agreement with theory[13-22].

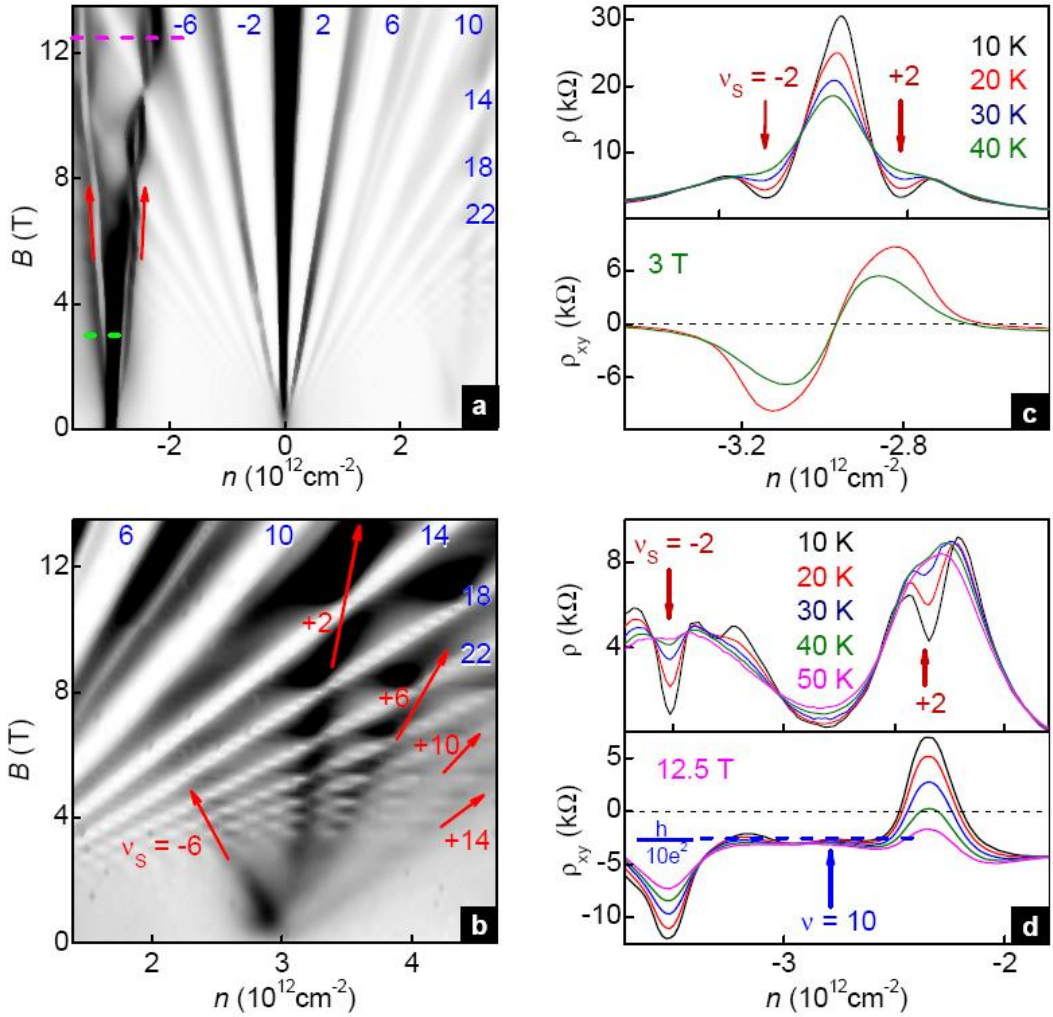

FIG. 2. **Quantization in graphene superlattices. a** – $\rho_{xx}(n,B)$ at 20 K. Grey scale: 0 (white) to 8.5 kOhm (black). **b** – Zoom-in near the electron-side secondary NP (grey scale: 0 to 1.1 kOhm). Blue numbers denote $\nu$ for the QHE states originating from the main DP. The red arrows in (a) mark the superlattice quantum states that evolve along $\nu_S = \pm 2$ (the arrows are shifted, not to obscure the white stripes). In (b), the red arrows indicate $\nu_S$ for the electron-side NP. It is difficult to decisively associate the electron-side superlattice LLs with any particular $\nu_S$, albeit the strongest peak in $\rho_{xx}$ evolves almost as $\nu_S = +2$. **c,d** – Detailed behavior near the hSNP in fields marked by the dashed lines in (a). Data are for device A but the quantization behavior was found universal for all the devices. An exception is the white stripes at $\nu_S = \pm 2$, which were often smeared by inhomogeneity so that only broader maxima in $\rho_{xx}$ remained (similar to the curves at 50K). Nonetheless, the narrow extrema in $\rho_{xy}$ associated with the minima in $\rho_{xx}$ (as seen in d), were always present.

Figure 2 shows evolution of $\rho_{xx}(n)$ with increasing perpendicular magnetic field $B$. Near the main DP, we observe the standard[30] quantum Hall effect (QHE) for graphene, with plateaus in $\rho_{xy}$ and zeros in $\rho_{xx}$ at filling factors $\nu \equiv n\phi_0/B = \pm 2, 6, 10, \ldots$ where $\phi_0$ is the flux quantum. Fan diagrams around the secondary DPs are different (Fig. 2). For the hSNP, its resistance peak first broadens with increasing $B$ and then splits into two maxima. The maxima follow superlattice's filling factors $\nu_S = \pm 2$, where the carrier density is counted from $-n_S$. In the middle of each maximum, there is a profound minimum (narrow white stripes in Fig. 2a). The minima in $\rho_{xx}$ are accompanied by positive and negative extrema in $\rho_{xy}$ (Fig. 2c-d). This shows that electron-like cyclotron trajectories in graphene's valence band persist into quantizing $B$. With decreasing $T$, $\rho_{xx}$ inside the narrow minima tends to zero and the corresponding extrema in $\rho_{xy}$ become increasingly more pronounced, the behavior characteristic for the development of Shubnikov de Haas oscillations into QHE states (Fig. 2c-d). The $T$ dependence yields a QHE gap of $\approx 20$meV (Supplementary Info, Fig. S6). Unlike cyclotron gaps, this one is practically independent of $B$ as seen also from the fact that the white stripes in Fig. 2a do not widen. With increasing $T$, the QHE states at $\nu_S = \pm 2$ wash out but the maxima in $\rho_{xx}$ persist up to 150K.

Another important feature of the observed fan diagrams is multiple peaks in $\rho_{xx}$ accompanied by zeros or deep minima in $\rho_{xy}$. This is seen most clearly for devices where doping sufficiently higher than $n_S$ can be achieved (Fig. 3). Furthermore, in all our devices near the hSNP, $\rho_{xy}$ repeatedly changes its sign with increasing $B$, indicating recurrent appearance and disappearance of electron-like orbits within graphene's valence band (Figs. 3b,3e and Figs. S4-S5 of Supplementary Info). This means that, for a given $n$, the magnetic field alone can repeatedly generate new NPs. The "third-generation" NPs are periodic as a function of $n$ and $1/B$ and make into distinct groups aligning parallel to the $n$-axis (Fig. 3 and section 3 of Supplementary Info). Their periodicity in $1/B$ is accurately described by unit fractions $\phi_0/q$ of the magnetic flux $\Phi = B \times S_\otimes$ per superlattice unit cell area $S_\otimes$ where $q$ is integer. In the conduction band, the fan diagrams also exhibit LLs fanning from the secondary DP, and numerous third-generation NPs with the same $1/B$ periodicity are clearly seen in Figs. 2b,3a,3b. These features are weaker than those in the valence band. For example, the resistivity peak at the electron-side NP fades away already in $B \approx 1$T (Fig. 2b) and we did not observe the secondary QHE in the conduction band.

The observed superlattice behavior points at complex spectral changes induced by quantizing $B$. From a theory point of view, the problem is somewhat similar to that originally discussed by Zak[23] and Hofstadter[9] and later considered for 2D electrons[3] in semiconductor superlattices and for Dirac fermions in twisted bilayers[24]. The most general but not strictly proven prediction is that superlattice spectra should be 'self-similar', that is, consist of multiple clones of an original spectrum, which appear at such $B$ that $\Phi = \phi_0(p/q)$ where both $p$ and $q$ are integer. Our case of graphene-on-hBN is analyzed in Supplementary Info, and the main theory results are summarized in Figs. 3c-d. Figure 3c shows that the superlattice potential leads to an additional structure within each LL, and this effectively broadens them with increasing $B$. The structured LLs originating from the main and secondary DPs strongly mix at high doping $|E| \geq E_S$. The resulting pattern is different from that in semiconductor superlattices with a parabolic spectrum and weak modulation[3,7,8]. In the latter case, the fractal structure within each LL can be described by the original Hofstadter butterfly[9] that appears periodically as a function of $\phi_0/\Phi$. In our case of the Dirac-like spectrum and strong modulation, the fractal pattern depends on the LL index $N$ and $B$ (section 5 of Supplementary Info).

The calculated spectrum allows us to understand many features observed experimentally. Indeed, Fig. 3c exhibits a clear self-similarity such that magnetic states tend to entwine at $\Phi = \phi_0 p/q$ reflecting the fractal structure of the pattern. The strongest entwining occurs for unit fractions (that is, $p \equiv 1$), and this results in an overall $1/B$ periodicity with a period of $S_\otimes/\phi_0$, in agreement with the experiment (Fig. 3a-b). The periodicity can be traced back to the fact that at $\Phi = \phi_0/q$ the superlattice problem can be mapped onto the case of a new superlattice that has a unit cell $q$ times larger than original and placed in zero effective field[23]. An example of the resulting magneto-electronic (Zak) bands[23] is given in Fig. S8 of Supplementary Info.

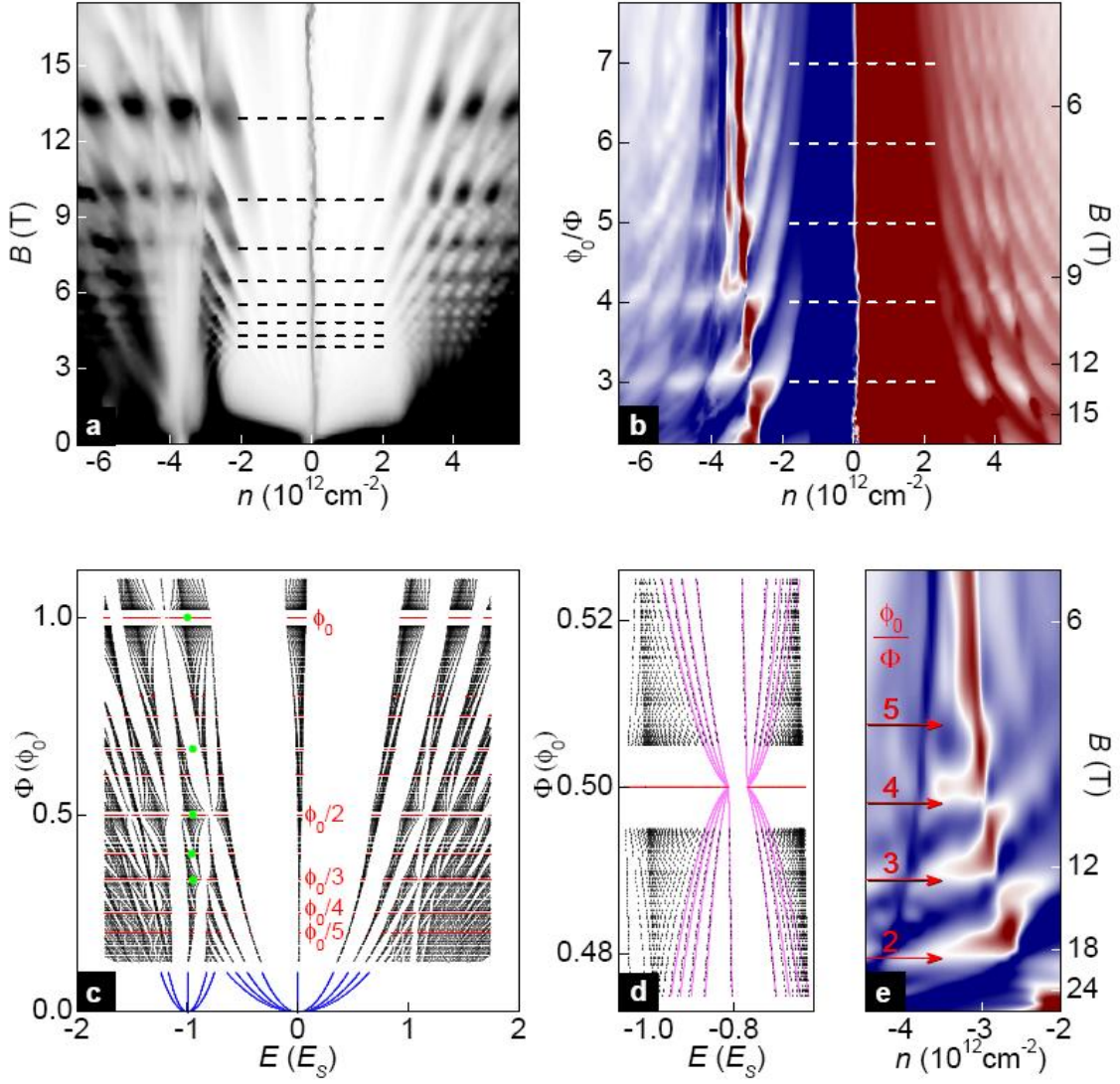

FIG. 3. **Zak's cloning of third-generation Dirac points. a** – Longitudinal conductivity $\sigma_{xx} =\rho_{xx}/(\rho_{xx}^2 + \rho_{xy}^2)$ as a function of $n$ and $B$ (device B). The dashed lines indicate $B_q = (1/q)\phi_0/S_\otimes$ with $q = 3$ to 10, where $S_\otimes$ is determined from the measured $n_S \approx 3.8\times10^{12}$cm$^{-2}$ as $S_\otimes = 4/n_S$ which corresponds to the complete filling of the first SBZ[22]. Grey scale: 0 (white) to 2.2 mS (black). **b** – $\rho_{xy}$ for the same device as a function of $n$ and $1/B$ with the left axis is in units of $\phi_0/B\times S_\otimes$. The white lines show Zak oscillations' periodicity. Color scale: navy to white to wine correspond to -2 to 0 to 2 kOhms. $T = 2$K for both plots. **c** – Hofstadter-like butterfly for graphene-on-hBN superlattice. The electronic states are calculated following the approach of ref. 24 and shown by black dots. For simple fractions $p/q$, we plot energies of the states in red. Regions at $\Phi$ near $\phi_0 p/q$ are left empty because of too large supercells required for the calculations[22,24]. The blue curves show low LLs in small $B$, which were calculated analytically for original and secondary Dirac fermions with parameters of the zero-$B$ spectrum. The green dots indicate the position of the Fermi level for $n = -n_S$ (Supplementary Info). **d** – Section of the butterfly in (c) with superimposed LLs calculated as a function of $\delta B$ (magenta). **e** – $\rho_{xy}(n,1/B)$ for device C ($n_S \approx 3.6\times10^{12}$cm$^{-2}$) measured up to 29T. Left and right axes are as in (b). The same oscillatory behavior is found for all our devices and seen most clearly near the hSNP, where new NPs appear periodically at $\phi_0/\Phi = q$ as shown by the arrows. Navy to white to wine: -3 to 0 to 3 kOhm; $T = 20$K. Full diagrams $\rho_{xy}(n,B)$ for devices B and C are provided in section 3 of Supplementary Info.

We find that in our case Zak's bands feature slightly gapped Dirac spectra. This finding is illustrated in Fig. 3d which, as an example, magnifies a part of Fig. 3c near the hSNP at $\Phi = \phi_0/2$. By using the Zak spectrum calculated for effective zero filed $B_2 = \frac{1}{2}\phi_0/S_\otimes$ we then obtain its Landau quantization in small reduced fields $\delta B$ $= B - B_2$. The resulting LLs evolve as $\pm\sqrt{N|\delta B| + \Delta^2}$ and are shown in Fig. 3d ($\Delta$ is the gap for the local Dirac-like spectrum). One can see that away from the empty horizontal region inaccessible in the calculations (section 5 of Supplementary Info), the Landau quantization of Zak's spectrum in the effective $\delta B$ (magenta curves) yields practically the same electronic states as shown by black dots. Similar local Dirac spectra are found in other parts of the moiré butterfly[24] for all $B_q = \phi_0/qS_\otimes$, in agreement with the numerous third generation NPs seen in the experiment. By using the calculated spectra, we have also determined the occupancy of Zak's minibands for the case of 4 holes per moiré supercell (that is, at the hSNP) and found that for $\Phi=\phi_0/q$, the Fermi energy lies inside the corresponding Zak minibands, whereas for $\Phi=\phi_0/(q+½)$, it lies inside gaps. This explains the experimentally observed oscillations in $\sigma_{xx}(B)$.

Another experimental feature revealing Zak's minibands and the hierarchy of superlattice gaps is the prominent QHE gaps at $\nu_S =\pm 2$ near the hSNP (Fig. 2). At small $B$, they can be considered as a result of Landau quantization for secondary Dirac fermions and separate the zero LL from the rest of the spectrum (Fig. 3c). In higher $B$, these gaps saturate being limited in size by the presence of van Hove singularities at the edges of the SBZ (Fig. 1a). One can see in Fig. 3c that as Zak's minibands become increasingly more pronounced the secondary LLs intertwine with main LLs and, at high doping, become indistinguishable from the latter. Therefore, the complex pattern of LLs in Fig. 3 at high doping can no longer be interpreted in terms of Landau quantization of either main or secondary Dirac fermions. This becomes a "Hofstadter-Landau" butterfly, specific to our strong-modulation regime and the linear spectrum. The largest fractal gaps near the hole-side DP in Fig. 3c are in agreement with the $\nu_S =\pm 2$ QHE states observed experimentally, which exhibit activation energies almost independent of $B$. This behavior is different from the case of weak modulation in semiconductor superlattices[6-8] where LLs become structured but do not intertwine. However, if $B$ is increased to $\Phi > \phi_0$, this can drive graphene superlattices into the regime of weak modulation, too (Fig. S9 of Supplementary Info). This regime is outside the range of $B$ available in our experiment. In addition to the discussed "large minigaps", our experimental data also reveal reproducible small scale structure that cannot be traced back to either main or secondary NPs (see, e.g., Figs. 3a,b near the hSNP in high $B$; Fig. S5a of Supplementary Info). We attribute these fine features to further fractalization of the superlattice spectrum so that isolated LLs for the third-generation Dirac spectra start being resolved (Fig. 3d). This is similar to the intra-LL features reported in semiconductor superlattices[7,8] and warrants further investigation.

To conclude, graphene superlattices can be reliably fabricated for various types of transport measurements. This opens new lines of enquiry, particularly in quantizing $B$ where such a rich behavior has already been revealed that its full understanding requires much further work, both theoretical and experimental. The demonstrated possibility to create quasi-gaps at specifically chosen energies by controllably rotating graphene or other 2D crystals within van der Waals heterostructures[25] can be exploited to design novel electronic and optoelectronic devices.

**Methods**

Our devices were multiterminal Hall bars fabricated following the procedure described in ref. 27. In brief, monolayer graphene is deposited on top of a relatively thick (>30 nm) hBN crystal[28] and then covered with another hBN crystal. The encapsulation protects graphene from environment and allows high μ, little residual doping (<10$^{11}$ cm$^{-2}$) and little charge inhomogeneity[27]. The interfaces between graphene and hBN are atomically clean over the entire active device area[25]. The whole stack is assembled on top of an oxidized Si wafer which serves as a back gate. To align the crystal lattices, we used an optical microscope to choose straight edges of graphene and hBN crystallites, which indicate principal crystallographic directions (see, e.g., Fig. 2 of ref. 30). During the assembly, graphene was rotated relatively to the bottom hBN to make their edges parallel. We estimate our alignment accuracy as ≈1°. The top hBN was then rotated by ~15° with respect to the edges, which ensured no spectral reconstruction at $E_S$ <1eV due to the second graphene-hBN interface.

Gate dielectric's breakdown for oxidized silicon wafers in practice occurs in fields <0.4V/nm and this limits achievable $n$ to <7×10$^{12}$cm$^{-2}$ ($E_S$ <0.35eV). Accordingly, the observation of secondary DPs requires alignment with θ ≤1°(ref. 12). For random deposition of graphene on hBN, chances of finding transport devices exhibiting superlattice features are a few %, even if measuring up to high gate voltages, which are rarely used to avoid accidental breakdown. Previously, we investigated >25 graphene-on-hBN devices[27] and neither of them exhibited any sign of superlattice effects. This shows that careful alignment is essential for the observation of secondary Dirac spectra in transport measurements.

## Supplementary Information

### #1 Transport properties of secondary Dirac fermions

Near the main and secondary NPs, our devices exhibited surprisingly similar carrier mobilities µ (see the main text). They were within a range of ≈20–100×10³ cm²V⁻¹s⁻¹ depending on sample. No short-range resistivity term that often yields a sublinear dependence σ($n$) was noticeable in our devices.

As usual for graphene on hBN [S1-S2], near the main NP we find µ to be practically independent of $T$ within our entire $T$ range, which was limited to 150 K to avoid breakdown of the gate dielectric. Near the electron-side secondary NP (eSNP), µ also shows only a weak $T$ dependence. In stark contrast, there is a strong $T$ dependence near the hSNP (Fig. 1a of the main text) such that µ falls below 10,000 cm²V⁻¹s⁻¹ at 150K. The behavior did not change significantly below 10 K.

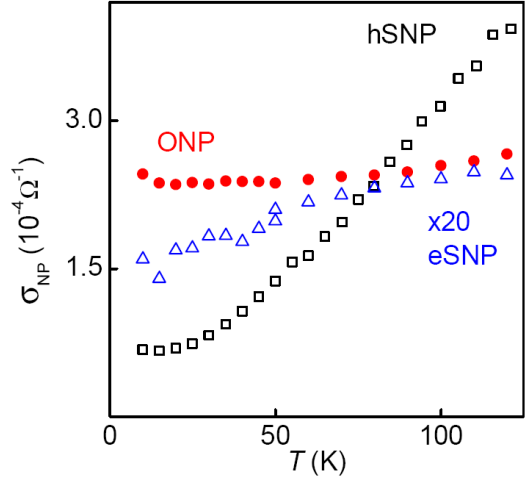

FIG. S1. $T$ dependences of minimum conductivity at the main and secondary NPs. For the electron-side NP, $\sigma_{NP}$ is scaled by a factor of 20.

Another notable difference between the three NPs is that they exhibit different $T$ dependences of their minimum conductivities $\sigma_{NP}$ (Fig. S1). For the hSNP, $\sigma_{NP}$ increases by a factor of 10 between liquid-helium $T$ and 150K. For the eSNP and main NP, changes in $\sigma_{NP}$ are small (<50%), similar to the standard behavior for graphene with similar µ [S1-S3]. Despite the strong $T$ dependence at the hSNP, it does not follow the activation behavior but evolves linearly with $T$ and then saturates below 20K (Fig. S1). We believe that this dependence is unlikely to be caused by a gap opening or localization effects because we have found $\sigma_{NP}$ insensitive to small magnetic fields $B$ <0.1T [S4]. Similar $\sigma_{NP}(T)$ were reported for high-µ suspended devices and attributed to a combined effect of thermally excited carriers and $T$-dependent scattering [S5-S6]. The observed $T$ dependences cast doubt that an hBN substrate can induce large energy gaps envisaged theoretically [S7-S8].

In general, the observed transport properties and, especially, different $T$ behavior for hole- and electron-side Dirac fermions are puzzling and remain to be understood.

### #2 Thermal broadening of secondary Dirac points

Another important difference between the main and secondary NPs is their different thermal broadening. At low $T$, the main DP is broadened by charge inhomogeneity δ$n$, which is ~10¹¹cm⁻² in our aligned devices. As expected for such δ$n$ [S5-S6], we observe little additional broadening at the main NP with increasing $T$ (Fig. 1a). In contrast, the hSNP becomes strongly and visibly broader with $T$ despite high δ$n$ (Figs. 1a and S2). This broadening can be analyzed in terms of the number Δ$n_T$ of thermally excited charge carriers [S5-S6]. If δ$n$ is relatively small (δ$n$ leads to residual broadening at low $T$), thermal carriers provide a dominant contribution to

σ(n) at the NPs. Accordingly, the peak in ρ$_{xx}$ becomes lower and broader with increasing T and its top gets rounder. The speed of this broadening as a function of T depends on the density of states (DoS) available for thermal excitations. It was shown theoretically and observed experimentally that Δn$_T$ varies as $T^2$ and T for the linear and parabolic spectra in graphene and its bilayer, respectively [S5-S6].

We have employed the same procedure as described in detail in ref. S6 to probe the DoS at the secondary DPs in our graphene superlattices. An example of this analysis is shown in Fig. S2 that plots the total number of carriers, Δn$_T$+δn, at the main and hole-side NPs for device A of the main text. The hSNP broadens >10 times faster than the main NP but both evolve as $T^2$. Because the peak at the hSNP is large and broadens rapidly, our experimental accuracy is high and the observed square T dependence unequivocally proves that the spectrum near the hSNP is linear, that is, Dirac-like. The eSNP also exhibits rapid thermal broadening but, for the small ρ$_{xx}$, quantitative analysis is difficult in this case.

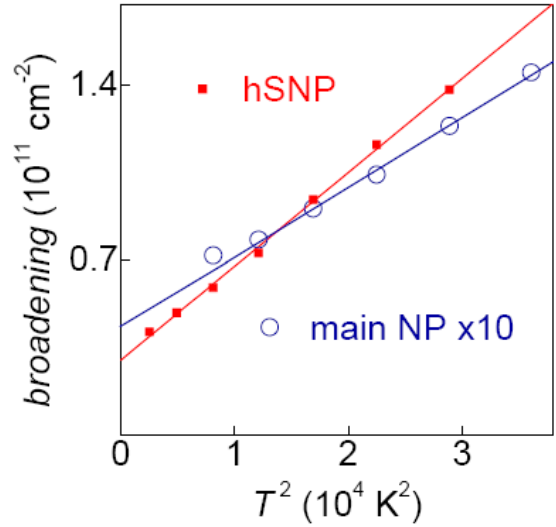

Fig. S2. Number of thermal charge carriers at the main and secondary NPs. The ratio between slopes of the red and blue lines is ≈13. The $T^2$ dependence proves that the spectrum at the new NPs is linear.

For a Dirac spectrum with degeneracy N, Δn$_T$ is proportional to $N/v_F^2$ [S6]. The average Fermi velocity $v_F^S$ for the secondary Dirac spectra in graphene on hBN was estimated as ≈0.5$v_F$ [S9], in agreement with theory [S10-S11]. Therefore, the observed Δn$_T$ ratio of 13±3 (Fig. S2) points at a triple degeneracy for the hole-side secondary DPs, consistent with the models that assume only a scalar potential modulation [S9-S13]. We also note that the main NP (blue curve) exhibits exactly the same speed Δn$_T$/$T^2$ of thermal broadening as previously reported for the NP in suspended graphene with little δn [S5], which shows good consistency of employing this approach for different graphene systems.

**#3 Further examples of Landau fan diagrams**
Figure S3 shows another superlattice fan diagram observed in our experiments. The central panel plots the entire diagram whereas the left and right panels zoom-in on the secondary NPs. In the conduction band, the third generation of DPs is seen an oscillatory network emerging beyond the eSNP, similar to the case in Fig. 2b and 3a of the main text. Near the hSNP, individual peaks in ρ$_{xx}$ due to third-generation DPs are not resolved as a function of n and merge into continuous bands, running parallel to the n-axis beyond the hSNP (see Figs. S3a and S4a). These bands can be referred to as Zak oscillations [S14] and are different from both Shubnikov-de Haas and Weiss oscillations.

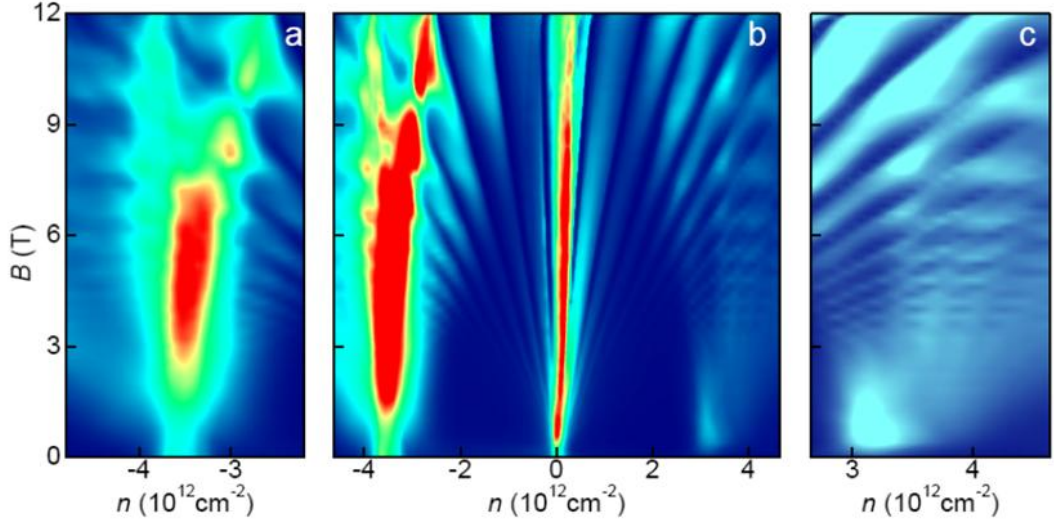

Fig. S3. Landau fan diagrams for device D. **b** – Complete diagrams $\rho_{xx}(n,B)$ showing the main and secondary NPs. **a, c** – Zooming in near the hole- and electron-side DPs, respectively. The blue-to-red scale is from 0 to 16, 8 and 1kOhm for plots a, b and c, respectively. The device exhibits somewhat higher charge inhomogeneity than device A of the main text and, accordingly, the hSNP is broader and its splitting occurs in higher $B$. The narrow minima in $\rho_{xx}$ along $\nu_S = \pm 2$ (such as in Fig. 2a of the main text) are not seen in this device, although the associated narrow extrema in $\rho_{xy}$ survive the inhomogeneity (see below). The data are taken by sweeping gate voltage at every 0.25T.

In $\rho_{xx}$ measurements, maxima due to third-generation DPs can be difficult to resolve as they often merge into continuous bands for a given $B$ (Fig. S4a). In this case, individual NPs are still seen clearly in Hall measurements. This is illustrated in Fig. S4a-b, which compares fan diagrams for $\rho_{xx}$ and $\rho_{xy}$ for the same range of $n$ and $B$. The Zak oscillations seen in $\rho_{xx}$ are split into separate spots in $\rho_{xy}$, similar to the case in Fig. 3a-b of the main text. The white spots in Fig. S4b correspond to deep minima in $\rho_{xy}$ and, near the hSNP, the Hall effect repeatedly changes its sign as a function of $B$. These minima are accompanied by maxima in $\rho_{xx}$. Zak oscillations as a function of $B$ are well described by unit fractions of $\phi_0$ per superlattice unit cell (Fig. S4c-d). This behavior is in good agreement with that reported for devices B and C in Fig. 3 of the main text and, in fact, was found in all our devices. For different devices, the observed $1/B$ periodicities varied according to their $S_\otimes$ determined from the same fan diagrams as $S_\otimes = 4n_S^{-1}$ [S11].

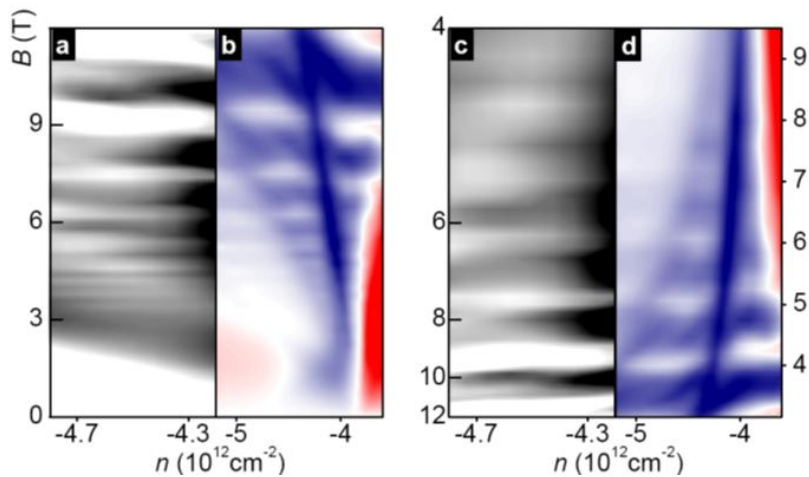

Fig. S4. Zak oscillations. **a** – $\rho_{xx}$ and **b** – $\rho_{xy}$ as a function of $n$ and $B$ beyond the hSNP. Device D as in Fig. S3. Grey scale in (a): 1.5 (white) to 2.8 kOhm (black). Color scale in (b): blue to white to red correspond to -0.2 to 0 to 0.2 kOhm. **c,d** – Same data replotted as a function of $1/B$. The left $y$-axis is in units of $B$; the right one in units $\phi_0/B \times S_\otimes$. It is clear that the oscillations are periodic in $1/B$ and correspond to unit fractions of $\phi_0$ per superlattice unit cells.

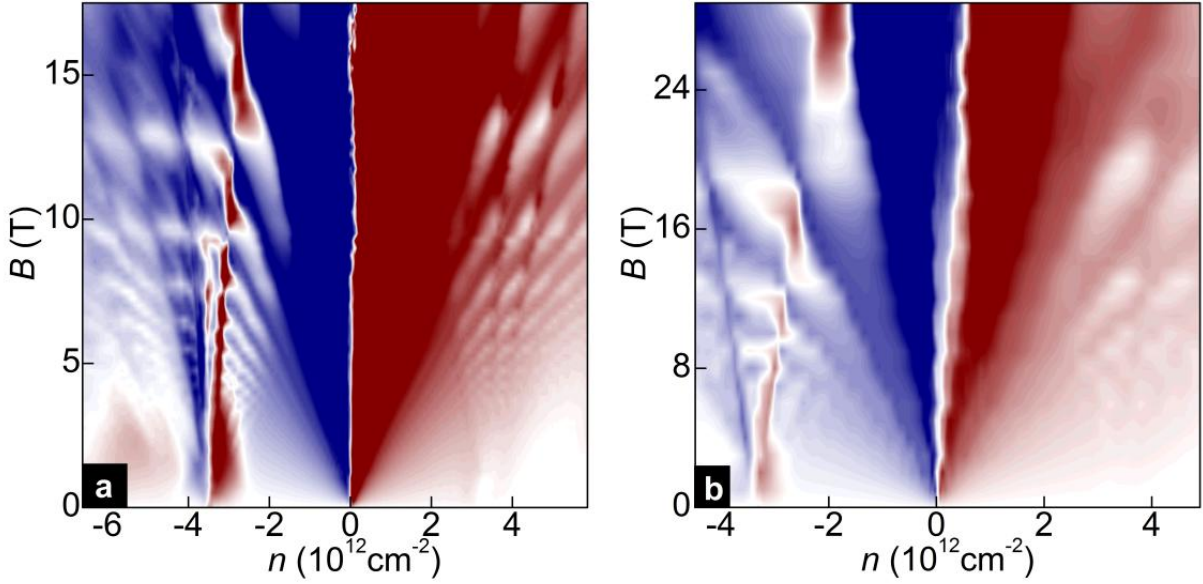

Fig. S5. High-$B$ behavior of $\rho_{xy}$ with numerous third-generation NPs. Scale: blue to white to red corresponds to -2 to 0 to 2 kOhm in (a) for device B and -6 to 0 to 6kOhm in (b) for device C. The data are taken by sweeping gate voltage at every 0.25T in (a) and 1T in (b) (this discreteness leads to the small-scale structure clearly visible at the lower-$B$ parts). The slight shift of the main NP in (b) is specific to this device and probably due to suppression of remnant doping by high $B$. Note that the oscillations near the eSNP do not lead to the sign change in the Hall effect but $\rho_{xy}$ still reaches very close to zero.

For completeness, Figure S5 shows the full Landau fan diagrams $\rho_{xy}(n,B)$ measured for devices B and C. The data partially appeared in Figs. 3b,e of the main text where the full diagrams were cropped and presented in a scale linear in $1/B$. Fig. S5 again shows repetitive reversals of the Hall effect with increasing $B$, a phenomenon that has never been observed in other systems.

#### #4 Superlattice QHE states

With reference to Fig. 2 of the main text, Figure S6 shows the QHE states running along $\nu_S = \pm 2$ at various $T$ in $B$ =5T, just before the central peak at the hSNP splits into two. The minima in $\rho_{xx}$ become deeper with decreasing $T$ (Fig. S6a) but do not reach the zero resistance state even at 1K, being blurred by charge inhomogeneity that suppresses the perfect edge state transport in our relatively narrow devices.

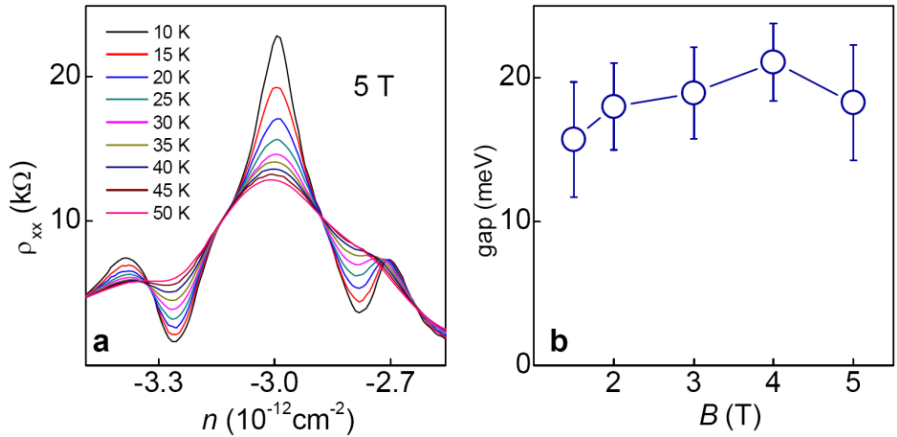

Fig. S6. Quantum Hall effect for secondary Dirac fermions. **a** – $T$ dependence near the hSNP in constant $B$. **b** – Corresponding energy gaps and their field dependence. The gaps were evaluated by analyzing $T$ dependences such as in (a) by using the Lifshitz-Kosevich formula (see, e.g., ref. [S3]). We did not investigate in detail the $T$ dependence after the central peak split in higher $B$ but, qualitatively, the gaps' size does not change up to 14T (see the $T$ dependence shown in Fig. 2d of the main text).

By analyzing *T* dependences such as in Fig S6a, we have obtained the corresponding gaps in different *B* (Fig. S6b). Within our experimental accuracy, the gaps for $\nu_S$ =-2 and +2 are equal and do not depend much on *B* (except for *B* where unit fractions of $\phi_0$ pierce the superlattice unit cell), consistent with the fact that the width of the narrow white stripes in Fig. 2a of the main text does not change.

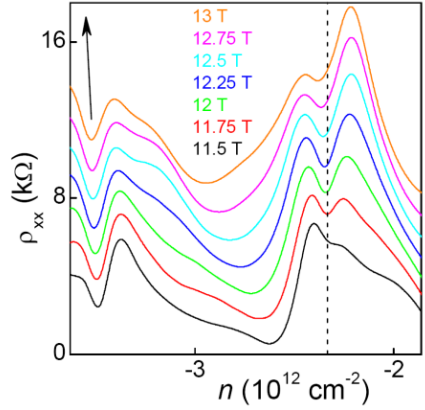

Fig. S7. Detailed evolution of the QHE states emerging near $\nu_S$ =±2 as a function of *B* as they cross the $\nu$ =-10 state originating from the main DP (Device A; *T*=20K). The curves are shifted vertically for clarity. The vertical line is to indicate little shift for the right ($\nu_S$ =+2) state. The arrow marks a fractal slope $\Delta\nu$ =-5/3.

Finally, we want to point out an intriguing behavior of the secondary QHE states running along $\nu_S$ =±2 when they 'hybridize' with the QHE states coming from the main NP. This is seen as the step-like waving of the white lines in Fig. 2a of the main text, which change their slopes each time the $\nu_S$ =±2 states cross the QHE states originating from the main DP. To examine this behavior further, Fig. S7 shows $\rho_{xx}$ in the interval where the $\nu_S$ =±2 states are intersected by the $\nu$ =-10 state. One can see that the position of the right minimum changes little with *B*. The changes (if any) are consistent with a small negative slope $\Delta n/\Delta B$ rather than running parallel to any positive $\nu_S$. The better developed minimum at $n \approx -3.5\times10^{12}$cm$^{-2}$ moves leftwards, as expected for this state that shows the general tendency to run along $\nu_S$ =-2 (Fig. 2a). However, the speed at which the minimum's position moves with *B* is lower than $\nu_S$ =-2 necessitates. Furthermore, Fig. 2d of the main text shows that, as $\rho_{xx}$ tends to zero, $\rho_{xy}$ develops symmetrically with respect to the Hall plateaus originating from the main NP. We speculate that, if this particular development continues, new QHE plateaus may appear at $h/e^2(1/\nu +1/\nu_S)$ where $h/e^2$ is the resistance quantum. For the case $\nu$ =-10 in Fig. 2d, this would infer $\rho_{xy}$ =-(3/5) and +(2/5)×$h/e^2$ and correspond to fractional fillings -5/3 and +5/2. To this end, we note that the best developed QHE state (deepest minimum in $\rho_{xx}$) runs parallel to $\nu$=-5/3 in the corresponding *B* interval as shown by the arrow in Fig. S7. The overall behavior may indicate that our fractal quantum Hall system supports a single-particle fractional QHE by mixing different integer QHE states.

**#5 Spectral characteristics of graphene superlattices**
As shown in Refs. S7-S12, there exist 3 principal scenarios for the superlattice spectrum of graphene placed on a hexagonal substrate. All these scenarios lead to secondary DPs at the edges of the lowest-energy moiré minibands in zero *B* [S11]. However, detailed spectra depend on size and relative strength of the phenomenologically introduced moiré pattern parameters, $u_{0,1,3}$ in the Dirac Hamiltonian for electrons in each of the two graphene valleys ($\varsigma = \pm 1$),

$$H_{moiré} = v_F \vec{p}\cdot\vec{\sigma} + u_0 f_1 + \varsigma u_3 f_2 \sigma_z + \varsigma u_1 (\vec{l}_z \times \nabla f_1)\cdot\vec{\sigma} \qquad (1)$$

where $\sigma_{x,y,z}$ are the Pauli matrices acting on the sublattice components of the electronic wavefunction [(A,B) in the valley K and (B,-A) in K'], $\varsigma = \pm 1$ for K and K' valleys; $f_1 = \sum_{n=0..5} e^{i\vec{b}_n\cdot\vec{r}}$ and $f_2 = \sum_{n=0..5}(-1)^n e^{i\vec{b}_n\cdot\vec{r}}$, where six vectors $\vec{b}_n$ (with $|\vec{b}_n| = b$) are obtained by consecutive 60° rotations; $\vec{b}_n = \hat{R}_{n\pi/3}\vec{G}$ of the reciprocal lattice vector $\vec{G} = [(1+\delta)\hat{R}_\theta - 1]\vec{b}_G$ of the moiré pattern, and $\vec{b}_G$ and $(1+\delta)\vec{b}_G$ are the principal reciprocal lattice vectors of aligned graphene and BN lattices, rotated by a small misalignment angle θ.

Three characteristic miniband spectra can be found [S11] for Dirac electrons described by the model in Eq. (1): (a) for small values of parameters $u_{0,1,3}$, strongly overlapping minibands without clearly separated band edges; (b) for particular relations between these parameters (for example, $u_{1,3} = 0$), a triplet of isolated secondary DPs with anisotropic Dirac velocities at the edge of the hexagonal mini Brillouin zone of moiré superlattice; and (c) more generically, one isolated secondary DP at a corner of the mini Brillouin zone (in each graphene valley) in either valence or conduction band, with a second-generation Dirac velocity of ≈0.5$v_F$. Except for special choice of moiré superlattice parameters, spectra of the Hamiltonian in Eq. (1) do not have electron-hole symmetry. Examples of the calculated characteristic miniband spectra for each of the three cases can be found in Ref. [S11].

Figures S8-S10 show examples of the magneto-electronic spectra expected for our graphene superlattices. In Fig. S8, we limit the plotted values to fluxes $0.1\phi_0 < \Phi < 0.6\phi_0$, which for our devices corresponds approximately to $B$ between 3 and 20T, that is, our typical experimental range. The calculated data are similar to those presented in Fig. 3c,d of the main text and obtained by using the procedure described in Ref. [S11]. We use 3 exemplary sets of moiré parameters, which are chosen to illustrate possible scenarios for graphene-on-hBN superlattices, taking into account the electron-hole asymmetry with a stronger secondary DP in graphene's valence band. Black dots in Fig. S8 present energies of states at the center of Zak's magnetic minibands found for arbitrary fractional flux values $\Phi = BS_\otimes = (2p/q)\phi_0$ [S14-S18]. In Figure S8a, we also show so-called spectral support [S16], that is, the entire miniband for several even and odd values of $q$ (blue intervals; $p$ =1).

In the lower-$B$ part of the plots, one can see remnants of the original Dirac spectrum with its Landau levels (LL) progressively broadened by the superlattice potential. To illustrate this fact, the red curves in Fig. S8a show several original LLs, in the absence of a superlattice potential. The superlattice spectra also contain reminiscence of Dirac-like quantized levels originating from secondary DPs. This is illustrated in Fig. S8a by another set of red curves beginning from ≈-0.5$v_F b$. These LLs evolve as $-0.51v_F b \pm 0.5v_F\sqrt{2N\hbar eB}$ with $N$ =0,1 , …. The green dots in Fig. S8a (also, Fig. 3c of the main text) show positions of the Fermi energy for $n = -n_S \equiv -4/S_\otimes$, that is, for the complete filling of the first moiré miniband in the valance band. These calculations are done by counting the number of filled magnetic bands (whose capacity and degeneracy depend on $p$ and $q$ [S14-S15]). The reason for us to focus on this particular density is that it corresponds to the half-filled zero LL originating from the secondary DP at the edge of the first moiré miniband in zero $B$. Therefore, this is the state that exhibits the initial (zero-$B$) change in the sign of Hall conductivity. Moreover, at $\Phi$ where this LL (zero $N$ for the secondary DP) splits into pairs of magnetic minibands, the Fermi level lies in a gap between them, which happens at $\Phi = \phi_0/(q+1/2)$ (for example, $\Phi$ =2/3, 2/5 or 2/7$\phi_0$). In this case, we also expect both $\sigma_{xx}$ to become zero and Hall conductivity to change sign. By counting filled states in magnetic minibands calculated for the flux $\Phi = \phi_0/q$, we find that in the latter case the Fermi level lies in the middle of Zak's magnetic bands and, although we have not find a way to determine the sign of $\rho_{xy}$ in this case, we can certainly state that Hall conductivity should once again change its sign and, therefore, take zero value somewhere in between two consecutive values of $\Phi = \phi_0/(q+1/2)$.

Furthermore, Fig. S8 shows that the zero-$N$ LL is robust and the superlattice potential broadens it relatively weakly over the entire range of $\Phi$ for this figure. This level is isolated from the rest of the Hofstadter spectrum by the large cyclotron gap $E_1$. For our superlattice modulation of ~50meV [S9], the $N = \pm1$ LLs are also reasonably well isolated. In contrast, LLs with higher $N$ strongly overlap, especially at concentrations near and above the secondary DPs. Therefore, graphene superlattices in quantizing $B$ of several tesla are typically in the regime of strong coupling [S19-S20]. Only for $\Phi > \phi_0$, the superlattices are expected to enter again in the regime of weak coupling where individual LLs are well isolated from each other, and the superlattice potential results in an internal structure within each Landau band [S19-S21]. It would require $B$ >30T to access this regime experimentally. For completeness, the corresponding spectra expected in such $B$ are shown in Fig. S9.

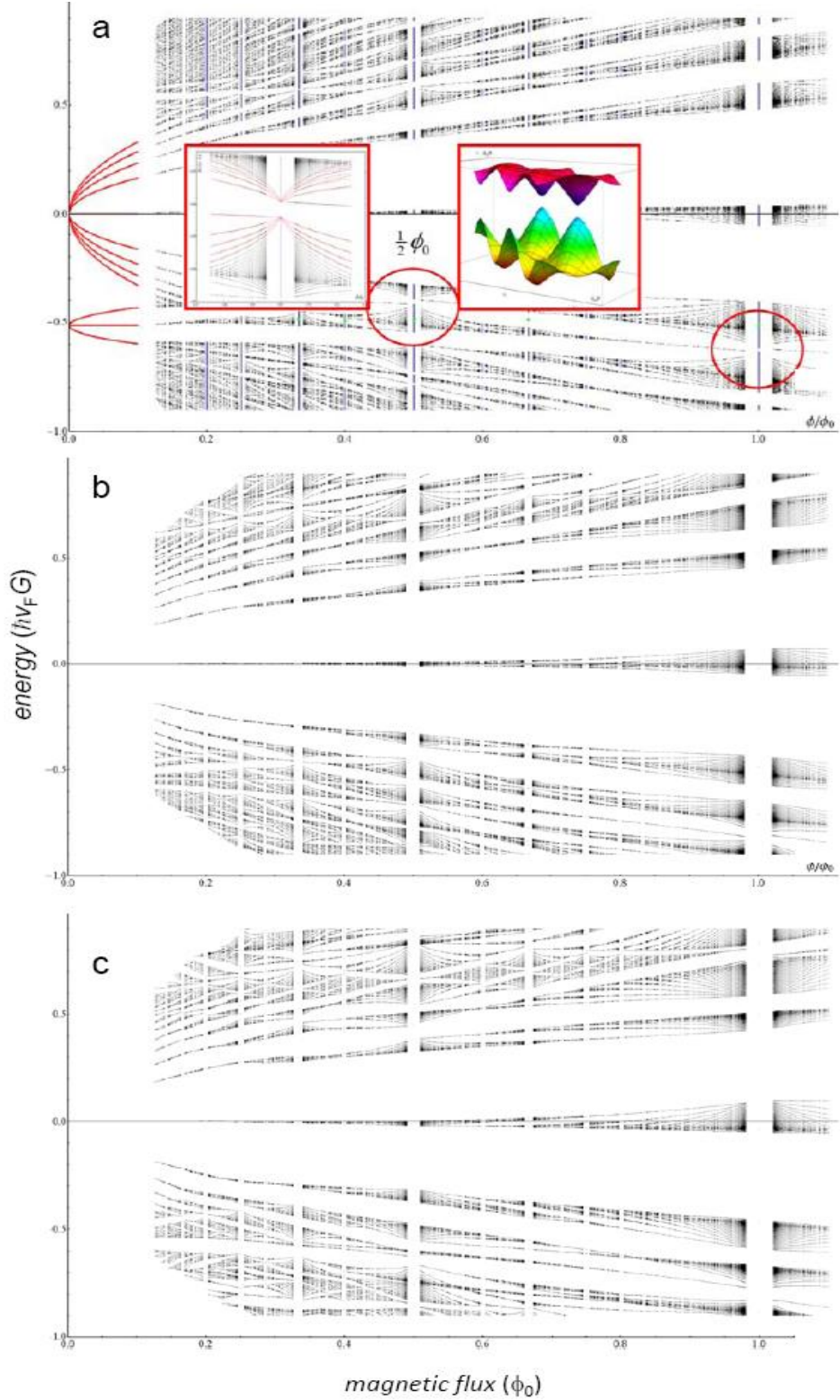

Fig. S8. Moiré butterflies spectra for characteristic superlattice potentials [S11]. **a** – $u_0$ = -0.031, $u_1$ = -0.015, $u_3$ = 0.054; **b** – $u_0$ = -0.072, $u_1$ = $u_3$ = 0.014 and **c** – $u_0$ = -0.1, $u_1$ = $u_3$ = 0 where $u$ are in units of $v_F b$. The energy scale is such that the secondary DPs appear at $\pm 1/2$. The right inset in (a) shows the energy dispersion (Zak's minibands [S14]) found in the energy range around the secondary DP for $\phi_0/\Phi$ =2; the left inset demonstrates that Zak's minibands are associated with a gapped Dirac-like spectrum and exhibit LLs characteristic of Dirac fermions (also, see ref. S11).

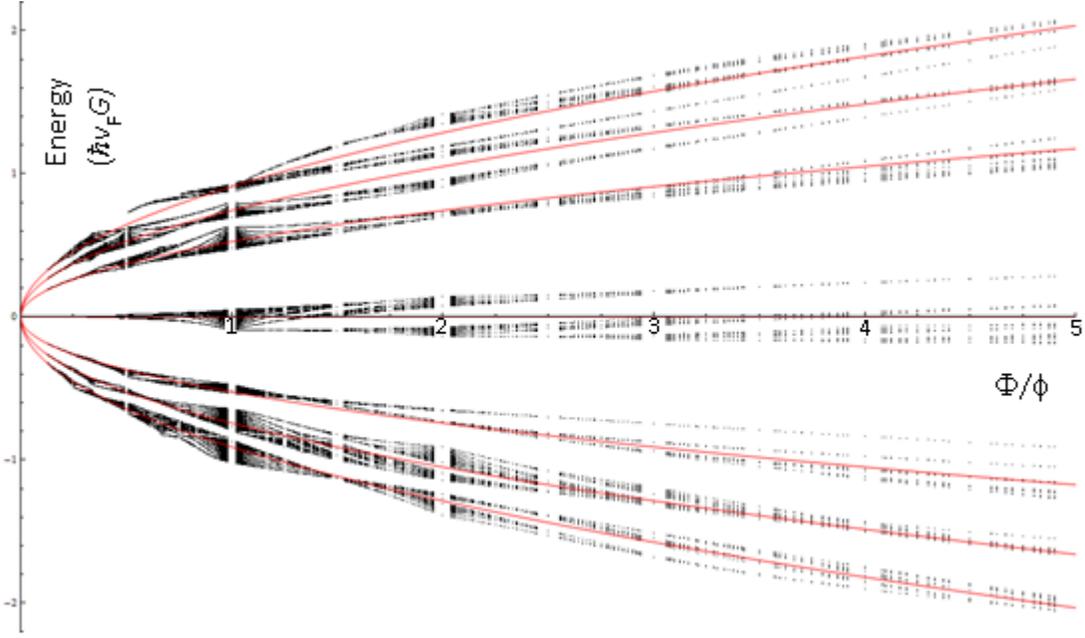

Fig. S9. Moiré butterfly for graphene superlattice in ultra-high *B*. Only the first LLs with $|N| \leq 3$ are shown. Superlattice potential *V* =60meV, that corresponds to $u_0$ = -0.1; $u_1$ = $u_3$ = 0. The original LLs (zero *V*) are shown in red. Black dots mark the superlattice states as in Fig. S8. The fractal structure with individual LLs (Hofstadter's butterfly) was previously studied for semiconductor superlattices in refs. [S19-S20]. An intra-LL structure is also noticeable in our Fig. 3a-b. However, the spectrum becomes particularly rich in the regime of strong coupling where the bands originating from different LLs overlap (Fig. S8). In our case, this condition is met for $\Phi < \phi_0$, that is, in *B* <30T.

The most striking feature of our moiré butterflies is self-similar sets of LLs that resemble those for Dirac fermions and repetitively appear over the entire superlattice spectrum (two circles in Fig. S8a point at characteristic regions). To understand the origin of these local quantized spectra, we have analyzed the miniband dispersion at fractional flux values $\phi_0/q$ and found that edges between pairs of consecutive minibands systematically display spectra $\omega_N \pm (u^2 k^2 + \Delta^2)^{1/2}$, that is, correspond to gapped Dirac fermions. One such dispersion is shown as an inset in Fig. S8a.

If we treat $\delta B = (B - B_q)$ as an effective magnetic field acting on electrons in Zak's magnetic minibands that appear at $B_q = (1/q)\phi_0/S_\otimes$ [S14], the gapped Dirac fermions give rise to a Landau-level fan with $E_{N=1,2,\ldots} = \varepsilon_N \pm \sqrt{Nu^2 e|\delta B| + \Delta^2}$ and $E_0 = \varepsilon_q + \Delta\, sign(\delta B)$. Using $\varepsilon_q(B) \approx \omega_N + c\delta B$ which takes into account an overall average shift of the parent Landau level, we have computed the corresponding spectrum and plotted it in the second inset in Fig. S8a (also, see Fig. 3d of the main text).

Finally, we replot one of our moiré butterflies (Fig. S8a) as a function of $\phi_0/\Phi$ (that is, $1/B$) and the energy renormalized to the energy $E_1$ of the 1st LL in the main spectrum. This is shown in Fig. S10 and allows easier comparison with the corresponding experimental plots in Figs. 3b,3e,S4c-d. The internal structure of LLs also becomes clearer in this presentation. One can see that the fractal spectra are different from the Hofstadter butterfly described by Harper's equation [S16] as well as from the moiré butterfly expected in twisted graphene bilayers [S17]. Moreover, there is no recurrence of the same fractal pattern within each Landau band. Such repetition of the Hofstadter butterfly is characteristic of semiconductor superlattices where a perfect periodicity within isolated Landau bands is expected for each unit between $\phi_0/\Phi = q$ and $q$ +1 [S19-S21]. In our case, we notice a different periodicity: *q*-th unit of *N*-th Landau level closely resembles (*q*+1)-th unit for (*N*+1)-th LL (see Fig. S10). Further work is required to understand fractality and properties of the intra-LL structure in graphene superlattices even in the limit of weak coupling.

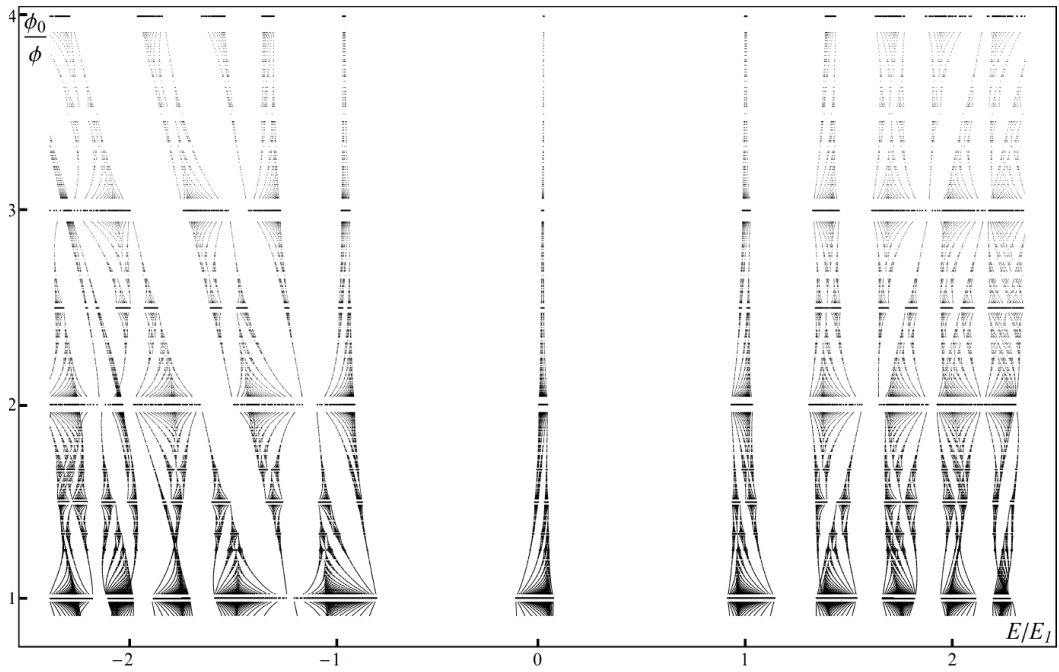

Fig. S10. Hidden periodicity of moiré butterflies. The superlattice spectrum in Fig. 3c of the main text is replotted as a function of $\phi_0/\Phi$ with the energy scale renormalized to $E_1$. It is the presentation standard for research on semiconductor superlattices [S19-21]. There are obvious Zak's oscillations with the energy gaps tending to close at integer $\phi_0/\Phi$. There is no obvious periodicity within each Landau band that occurs in semiconductor superlattices [S19-S21]. Nevertheless, notice that the first pattern ($\phi_0/\Phi$ between 1 and 2) within, for example, the 1st LL is similar to the second pattern for the 2nd LL, and so on. This periodicity involving both $\phi_0/\Phi$ and $N$ also survives in part for the hole side of the spectrum where the mixing between different LLs is much stronger.

## Supplementary references